\documentclass[aps,pra,twocolumn,superscriptaddress,notitlepage,groupedaddress]{revtex4-1}

\usepackage{graphicx}
\usepackage{mathptmx, textcomp}
\usepackage[usenames]{xcolor}
\usepackage{dcolumn}
\usepackage{bm}
\usepackage{amsmath}
\usepackage[colorlinks,urlcolor=blue,citecolor=blue,linkcolor=blue]{hyperref}
\usepackage{cleveref}
\usepackage{siunitx}
\usepackage{physics}
\usepackage{verbatim}
\usepackage{amsmath}
\usepackage[normalem]{ulem}

\def\te{{\rm e}}

\def\bk{{\bf k}}
\def\bp{{\bf p}}
\def\bq{{\bf q}}
\def\br{{\bf r}}

\def\calU{\mathcal{U}}

\def\T{{\mathcal T}}

\def\nn{\nonumber}

\def\TB{ {\mathcal{T}_{\rm B} } }

\def\aB{{a_{\text{B}}}}
\def\nB{{n_{\text{B}}}}

\def\Re{{ \rm Re }}

\def\BEC{{ \rm BEC }}

\def\ket#1{\left|#1\right\rangle}
\def\39K{$^{39}$K}
\def\87Rb{$^{87}$Rb}

\begin{document}

\title{Non-equilibrium quantum dynamics and formation of the Bose polaron}

\author{Magnus G. Skou$^1$}
\author{Thomas G. Skov$^1$}
\author{Nils B. J\o rgensen$^1$}
\author{Kristian K. Nielsen$^1$}
\author{Arturo Camacho-Guardian$^1$}
\author{Thomas Pohl$^1$}
\author{Georg M. Bruun$^{1,2}$}
\author{Jan J. Arlt$^1$}
\affiliation{$^1$ Center for Complex Quantum Systems, Department of Physics and Astronomy, Aarhus University, Ny Munkegade 120, DK-8000 Aarhus C, Denmark.}
\affiliation{$^2$ Shenzhen Institute for Quantum Science and Engineering and Department of Physics, Southern University of Science and Technology, Shenzhen 518055, China.}
\date{\today}

\maketitle
\textbf{Advancing our understanding of non-equilibrium phenomena in quantum many-body systems remains among the greatest challenges in physics. 
Here, we report on the experimental observation of a paradigmatic many-body problem, namely the non-equilibrium dynamics of a quantum impurity immersed in a bosonic environment~\cite{Landau1933,Pekar1946}. We use an interferometric technique to prepare coherent superposition states of atoms in a Bose-Einstein condensate with a small impurity-state component, and monitor the evolution of such quantum superpositions into polaronic quasiparticles. These results offer a systematic picture of polaron formation~\cite{Shashi2014,Shchadilova2016,nielsen2019,Mistakidis2019,Drescher2020} from weak to strong impurity interactions. They reveal three distinct regimes of evolution with dynamical transitions that provide a link between few-body processes and many-body dynamics. Our measurements reveal universal dynamical behavior in interacting many-body systems and demonstrate new pathways to study non-equilibrium quantum phenomena.}


Landau's quasiparticle theory~\cite{Landau1933} represents one of the most powerful concepts to understand many-body phenomena.  Originally the theory was developed to describe the interaction of an electron with phonons in a solid, leading to the formation of a quasiparticle~\cite{Pekar1946}. Nowadays it is widely used in many areas of physics, and forms the basis for understanding fundamental phenomena such as transport processes, colossal magnetoresistance~\cite{Mannella2005}, and superconductivity~\cite{Lee2006}. Yet, the dynamical processes leading to the formation of quasiparticles  has remained elusive in condensed matter systems due to their high densities and consequently fast evolution times. Ultracold quantum gases offer a unique quantum simulation platform~\cite{Bloch2012} to address this problem, as they permit the controlled generation of impurity atoms inside a fermionic~\cite{Schirotzek2009,Kohstall2012,Koschorreck2012,Massignan2014,cetina2016,Scazza2017,Schmidt2018,Yan2019,Darkwah2019} or bosonic~\cite{jorgensen2016,hu2016,Yan2020,ardila2019} quantum gas, where they perturb the surrounding medium to form quasiparticles, called polarons. The study of Bose polarons is particularly important because the linear sound dispersion of the BEC is analogous to that of phonons in crystals, prompting recent theoretical efforts to describe their non-equilibrium evolution~\cite{Shashi2014,Shchadilova2016,nielsen2019,Mistakidis2019,Drescher2020}. 
 
\begin{figure}[t!]
	\centering
	\includegraphics[width=1\columnwidth]{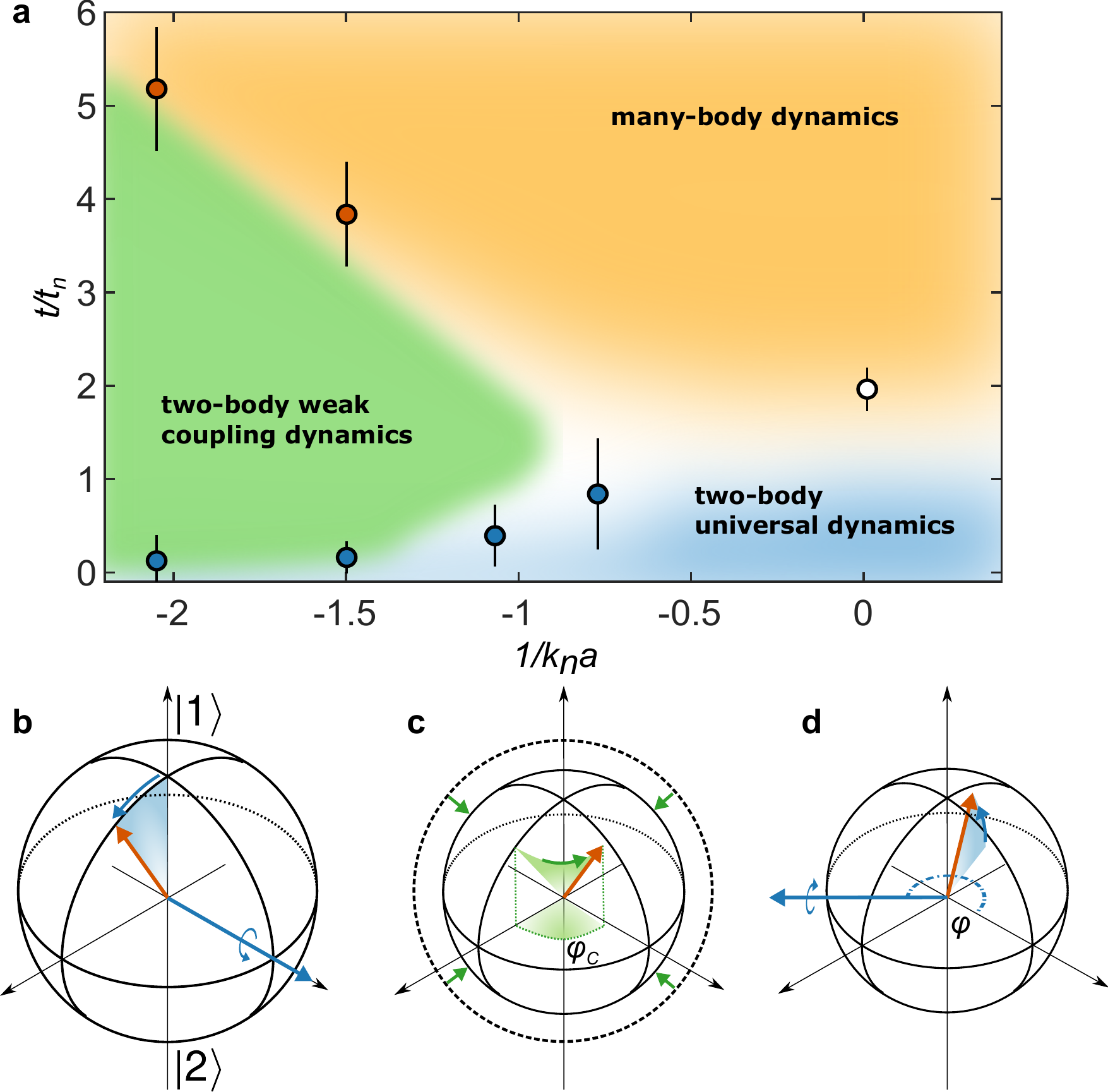}
	\caption{
		\textbf{Dynamical regimes of impurity evolution and experimental method.} \textbf{a,} Characteristic dynamical  regimes of impurity evolution as a function of the inverse interaction strength $1/k_n a$ and evolution time $t/t_n$. The measurements (circles) and theoretical analysis (colored areas) reveal three distinct dynamical regimes that extend from ultrafast two-body processes to the many-body regime of polaron formation. The blurring indicates smooth temporal transitions and the error bars correspond to the $1\sigma$ confidence intervals of the fitted values (blue) and the data resolution (red and white)~(Supplementary Information). \textbf{b}-\textbf{d,} Interferometric sequence to probe the dynamics illustrated using the collective spin of the atoms on the Bloch sphere. The north pole represents the initial state $|1 \rangle$ of the Bose-Einstein condensate and the south pole represents the impurity state $|2\rangle$. \textbf{b,} A short radio-frequency pulse prepares the system in a population-imbalanced collective superposition state. \textbf{c,} The subsequent evolution due to the interaction between the impurity state and its bosonic environment gives rise to a phase evolution $\varphi_\text{C}$ and a contraction of the Bloch sphere. \textbf{d,} A second pulse with variable phase $\varphi$ rotates the Bloch vector again whereafter the atomic spin population is obtained using an absorptive imaging technique. 
		}
	\label{fig:introductory}
\end{figure}

\begin{figure*}[t!]
	\begin{minipage}{\textwidth}
	\begin{center}
		\includegraphics[width=\textwidth]{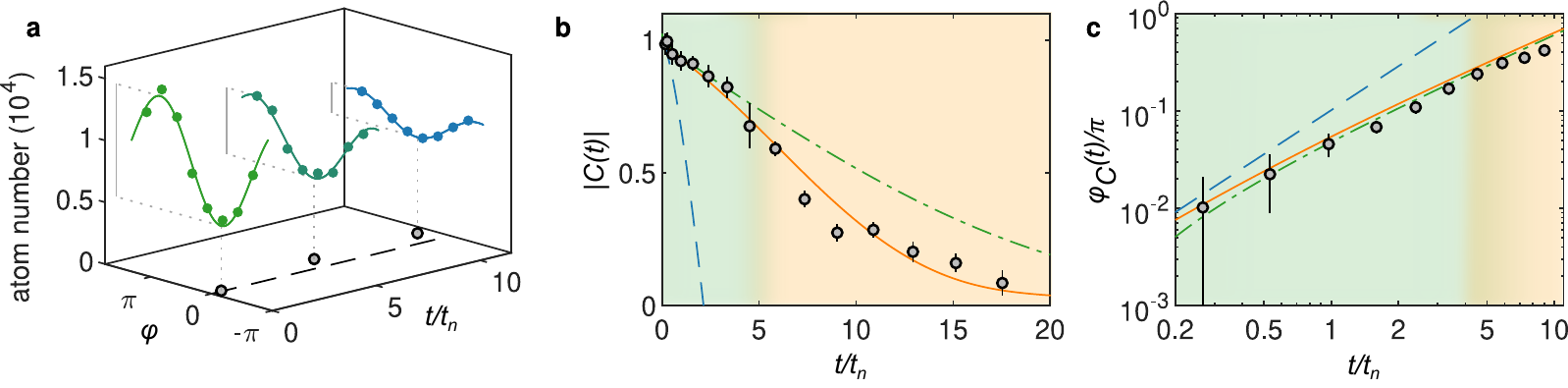}
	\end{center}
	\caption[d]{
		\textbf{Impurity dynamics at weak interaction strength.} \textbf{a,} Interference signal recorded at different evolution times as a function of the probe pulse phase $\varphi$ for an interaction strength $1/k_n a = -2$. Sinusoidal fits are shown as solid lines and the obtained amplitude and phase are indicated using gray lines and open circles respectively. \textbf{b,} Coherence amplitude $|C(t)|$. The impurity state decoheres due to interactions with the condensate. \textbf{c,} Phase of the coherence $\varphi_C(t)$. The impurity phase increases as the state rotates on the Bloch sphere, which at weak interaction strength is primarily due to the impurity state mean field interactions $E_{\rm mf}$. The dashed blue line shows the two-body universal $t^{3/2}$ and the dash-dotted green line shows the weak coupling $t^{1/2}$ prediction according to Eq.~\eqref{eq:Coherence_2body}. The solid orange line provides the diagrammatic description and the colored areas illustrate the theoretically predicted dynamical regimes from Fig.~\ref{fig:introductory}a. The error bars are $1\sigma$ confidence intervals of the fitted values.}
	\label{fig:measurement}
\end{minipage} 
\end{figure*} 

Here, we make use of this capability to induce and trace the non-equilibrium dynamics of a quantum impurity from its initial creation to the eventual formation of the Bose polaron. We drive an atomic transition to coherently create a small population of an impurity state in a Bose-Einstein condensate (BEC). Its interaction with the surrounding BEC induces fast quantum evolution, which we probe by monitoring the coherence between the initial state $|1\rangle$ of the atoms and the impurity state $|2\rangle$ interferometrically. This, in turn, yields a direct measurement of the time dependent Green's function of the impurity and thereby allows us to observe the non-equilibrium dynamics of the impurity that leads to the eventual formation Bose polarons in a BEC.

Our measurements reveal distinct regimes of impurity evolution and thus yield a complete map of its dynamical behavior, as shown in Fig.~\ref{fig:introductory}a. At short times, we observe a universal $\sim t^{3/2}$ decay of the impurity coherence~\cite{Parish2016} which does not depend on the coupling to the bosonic environment. This behavior originates in high-energy two-body scattering with the surrounding condensate and governs the initial relaxation. Thus it provides a clear experimental signature for such unitarity-limited processes. For weak interactions an intermediate dynamical regime emerges subsequently. Here low-energy collisions dominate the dynamical evolution which give rise to a distinct $\sim t^{1/2}$ decay of the impurity coherence.  At longer times, we eventually observe pronounced deviations from such power-law behavior, reflecting the emergence of many-body correlations that usher in the formation of the Bose polaron. The transitions between these dynamical regimes are shown in Fig.~\ref{fig:introductory}a. We observe remarkable agreement between theory and experiment for all  impurity interaction strengths and evolution times, which provides a quantitative understanding of the non-equilibrium dynamics of this quantum many-body system.

    
The experiment is performed with Bose-Einstein condensates of $^{39}$K atoms in the $|F = 1, m_F = -1 \rangle \equiv |1\rangle$ hyperfine ground state \cite{wacker2015}. The average condensate density  $n_{\rm B}$ sets the interaction independent energy scale $E_n = {\hbar}^2  (6\pi^2n_{\rm B})^{2/3} /2m$ of the system and the corresponding time scale $t_n = \hbar/E_n=\SI{4.8}{\micro s}$. For the controlled population of the impurity state we use a radio-frequency (rf) pulse to drive the transition to the $|F=1, m_F = 0 \rangle \equiv |2\rangle$ state~\cite{jorgensen2016}. The strength of the interaction is characterized by the dimensionless parameter $1/k_n a$, where $a$ is the scattering length for collisions between the impurity and the condensate state, and $k_n=(6\pi^2n_{\rm B})^{1/3}$ is the the characteristic wave number. We tune the scattering length $a$ by applying a homogeneous magnetic field in the vicinity of a Feshbach resonance at $\SI{114}{G}$~\cite{lysebo2010, jorgensen2016, tanzi2018}, which does not affect the scattering length $a_{\rm B}$ for collisions between the condensate atoms.

The interferometric sequence to populate the impurity state and probe its dynamics is illustrated in Fig.~\ref{fig:introductory} showing the evolution of the collective spin on the Bloch sphere~\cite{cetina2015, cetina2016, fletcher2017}. In this Ramsey-type scheme, we retain the orientation of the Bloch vector close to the initial one, corresponding to a low population of the impurity state. This allows the use of short rf-pulses which can resolve the evolution at times much shorter than $t_n$.

The measurement is initiated by applying a rf-pulse tuned to the atomic resonance with a duration of $\SI{0.5}{\micro s}$, well below the typical time scales of the subsequent impurity dynamics. As illustrated in Fig.~\ref{fig:introductory}b, this coherently generates an admixture of the impurity state with a small population $\sim 5\%$. Subsequently, this state evolves for a chosen time $t$ driven by the interaction between the impurity state and the surrounding condensate. Initially, this can be visualized as a rotation and shrinking of the Bloch vector as shown Fig.~\ref{fig:introductory}c. The corresponding dynamics of the impurity state is monitored by closing the interferometric sequence with a second rf-pulse with variable phase $\varphi$. As shown Fig.~\ref{fig:introductory}d, this second pulse implements a rotation of the Bloch vector around an axis defined by $\varphi$. The final spin population is obtained by measuring  three body recombination losses after a $\SI{2}{ms}$ relaxation time with absorption imaging. This interferometric sequence results in a sinusoidal dependence of the final atom number $N$ on the probe phase $\varphi$, as shown in 
Fig.~\ref{fig:measurement}a for various evolution times. We perform a fit $N(\varphi) = N_0 - \mathcal{A}\cos(\varphi-\varphi_C)$ for each evolution time $t$. Generally, amplitudes can  be extracted for longer evolution times even at small signal-to-noise ratio, where the phase determination fails.

Based on these fits we obtain the normalized coherence function $C(t)=|\mathcal{A}(t)/\mathcal{A}(0)|e^{i\varphi_C (t)}$~(Supplementary Information). This is in turn directly  proportional to the impurity Green's function $G(t)=-iC(t) = -i\bra{\psi_\text{BEC}}\hat{c}(t)\hat{c}^\dagger(0)\ket{\psi_\text{BEC}}$, where $|\psi_\text{BEC}\rangle$ describes the state of the BEC before the first rf-pulse and $\hat{c}^\dagger$ is the operator that creates an impurity in the condensate.
Consequently, $C(t)$ is directly related to the spectral function of the impurity, which we compute using both a two-body and a many-body description to obtain $C(t)$ throughout the impurity dynamics.

The initial dynamics can be calculated exactly for high energies where it is determined by two-body physics~\cite{braaten2010}. A Fourier transform gives the corresponding exact short time dynamics, which has the limiting forms~(Supplementary Information)

\begin{align}
C(t)  = &\begin{cases}
1  - (1 - i) \frac{16}{9 \pi^{3/2}} \left(\frac{t}{t_n}\right)^{3/2}& t\ll  t_a\\
 1 +\frac{2}{3\pi}(k_n|a|)^3 - iE_{\rm mf}t/ \hbar - (1 + i) \left(\frac{t}{t_{\rm w}}\right)^{1/2} & t\gg  t_a
\end{cases}
\label{eq:Coherence_2body}
\end{align}
where $E_{\rm mf} = 4\pi\hbar^2n_{\rm B}a/m$ is the mean field energy due to impurity state interactions with the BEC and $t_a=ma^2/\hbar$. For times $t\ll  t_a$, Eq.~(\ref{eq:Coherence_2body}) describes universal dynamics where the coherence of the impurity state decays with a power-law exponent of $3/2$ on a time scale $t_n$ \emph{independent} of the (non-zero) interaction strength (Fig.~\ref{fig:introductory}a, blue area). This universal short-time relaxation directly reflects the unitarity-limited scattering cross section for short-range interactions, which does not depend on $a$ for collision energies greater than $\hbar^2/m a^2$. Hence, the time $t_a$ marks the crossover (Fig.~\ref{fig:introductory}a, blue to green transition) to a regime where the dynamics is governed by the mean field phase evolution $E_{\rm mf}t/\hbar$, and the coherence decays with a power-law exponent ${1/2}$ on an interaction strength dependent time scale $t_{\rm w} = m / 32\pi \hbar n_{\rm B}^2 a^4$ (Fig.~\ref{fig:introductory}a, green area). This behaviour arises from weak two-body collisions with a constant cross section $\sim a^2$~\cite{nielsen2019}. 

An intuitive understanding of the power-laws in Eq.~(\ref{eq:Coherence_2body}) can be gained from the cross section $\sigma(k)=4\pi a^2/[1+(ka)^2]$ assuming that the rate of decoherence is given by the collision rate $\dot{C}(t) \sim -n_\text{B}\sigma v$. At a time $t$ after initializing the system, decoherence is caused by coupling to states with $E\sim \hbar /t$ setting the wave number $k\sim \sqrt{m/\hbar t}$ and the collisional velocity $v\sim \sqrt{\hbar/mt}$. For $t \ll  t_a$ the cross section is unitarity-limited $\sigma \sim 1/k^2 \sim \hbar t/m$ and integrating the decoherence rate yields the universal limit $C(t)\sim(t/t_n)^{3/2}$. At longer times $t \gg  t_a$ the cross section is determined by low energy collisions $\sigma \sim a^2$ and integrating gives $C(t)\sim(t/t_\text{w})^{1/2}$ in accordance with the weak coupling limit~(Supplementary Information).

At later times, interactions between multiple particles lead to pronounced deviations from the two-body prediction given by Eq.~\eqref{eq:Coherence_2body} and the system enters a regime of many-body dynamics (Fig.~\ref{fig:introductory}a, orange area). We describe this many-body dynamics using a diagrammatic theory~(Supplementary Information), which has previously been applied to the equilibrium physics of Bose polarons~\cite{Rath2013,ardila2019}. Since our many-body theory  contains the dominant two-body processes, it moreover recovers the two-body prediction of Eq.~\eqref{eq:Coherence_2body} for short times. For weak interactions, deviations from two-body weak coupling  $t^{1/2}$ dynamics occur at times $\sim \hbar/E_{\text{mf}}$ and signal the onset of many-body physics (Fig.~\ref{fig:introductory}a, green to orange transition).
However, for large interaction strengths where $E_{\text{mf}} >  \hbar^2/ma^2$ and consequently $|1/k_n a|<(2/3\pi)^{1/3}$, the many-body dynamics emerges directly from the initial universal regime at times $\sim 1.4 t_n$ (Fig.~\ref{fig:introductory}, blue to orange transition). We emphasize that these changes in dynamical behavior correspond to smooth temporal crossovers, as indicated by the blurred boundaries in Fig.~\ref{fig:introductory}a.


\begin{figure}[tb]
	\centering
	\includegraphics[width=1\columnwidth]{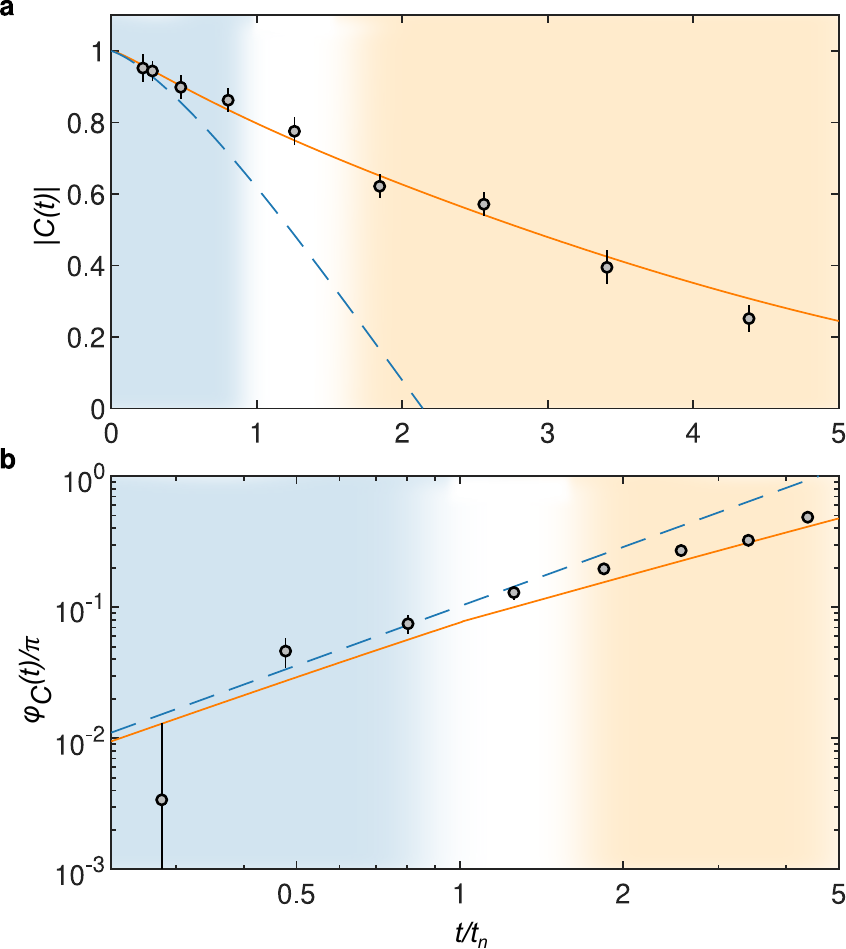}
	\caption{\textbf{Impurity dynamics at intermediate interaction strength.} \textbf{a,} Coherence amplitude and \textbf{b,} phase evolution
		for $1/k_n a = - 0.77$. The two-body universal $t^{3/2}$ prediction of Eq.~\eqref{eq:Coherence_2body} is shown as a dashed blue line. The solid orange line 
		is the diagrammatic prediction and the colored areas illustrate the theoretically predicted dynamical regimes from Fig.~\ref{fig:introductory}a. The error bars are $1\sigma$ confidence intervals of the fitted values.
	}
	
	\label{fig:exponent_and_timescales}
\end{figure}

Figure~\ref{fig:measurement}, b and c, show the measured coherence amplitude and phase in the regime of weak interactions. Both measured quantities agree well with the $t^{1/2}$ evolution given by Eq.~(\ref{eq:Coherence_2body}) for $t\lesssim \hbar/E_{\rm mf}$. The transition between the universal and weak coupling two-body dynamics at short times is extracted by simultaneously fitting the coherence amplitude and phase with the general two-body description~(Supplementary Information) using the transition time as a free parameter. These times are shown in Fig.~\ref{fig:introductory}a (blue data points). Moreover, the subsequent transition to the many-body regime is identified with the time when the observed coherence amplitude deviates more than two standard deviations from the prediction of Eq.~\eqref{eq:Coherence_2body}, also shown in Fig.~\ref{fig:introductory}a (red data points). Finally, we compare the observations to the diagrammatic prediction which captures the dynamics on all time scales. In particular, the many-body behavior is clearly visible for the amplitude at long evolution times. In principle, theory predicts that $|C(t)|$ decays towards the quasiparticle residue~\cite{Shchadilova2016,nielsen2019}, however, experimentally this is not observed due to additional decoherence processes~(Methods). The excellent agreement between theory and experiment nonetheless provides a benchmark for our measurement approach and theoretical understanding.

For intermediate interaction strengths the initial coherence amplitude and phase display universal $t^{3/2}$ dynamics extending for longer evolution times as illustrated in Fig.~\ref{fig:exponent_and_timescales}. There is no regime exhibiting $t^{1/2}$ dynamics since $t \gg t_a$ is not reached before the smooth transition to many-body dynamics. Furthermore, a measurement of the transition to the many-body regime is prohibited by the inhomogeneous density of our condensate, which obscures the observation of a clear deviation from Eq.~\eqref{eq:Coherence_2body}. Nonetheless, we obtain excellent agreement with the diagrammatic prediction for all evolution times which includes these experimental effects.

\begin{figure}[tb]
	\centering
	\includegraphics[width=1\columnwidth]{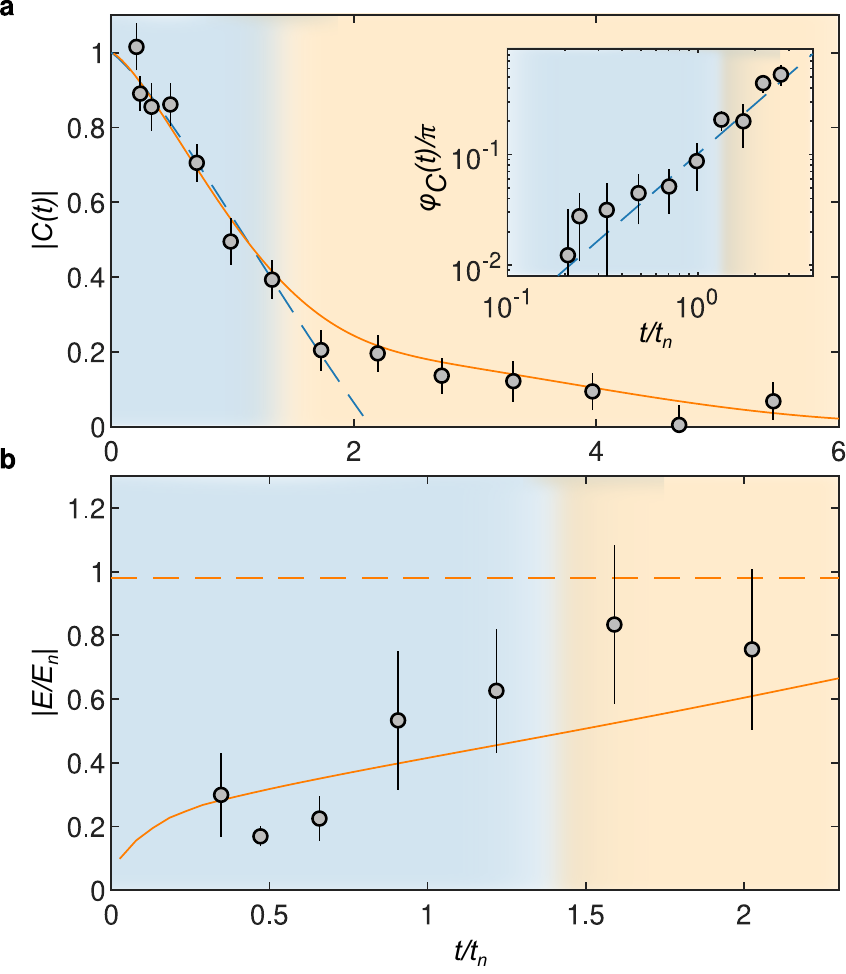}
	\caption{
		\textbf{Impurity dynamics at unitarity.} \textbf{a,} Coherence amplitude and phase evolution (inset). The fast initial decay of the coherence amplitude is in good agreement with the unitary two-body prediction of Eq.~\eqref{eq:Coherence_2body} shown as a dashed blue line. At longer times many-body physics dominates the decay, which is well described by a diagrammatic description that accounts for many-body effects shown as a solid orange line. \textbf{b,} Instantaneous energy obtained from the time derivative of the observed phase. The measured energies agree with the result of the diagrammatic theory shown as a solid orange line and approach the expected equilibrium energy of the Bose polaron $E_\text{p}$ marked with a dashed line. The colored areas illustrate the theoretically predicted dynamical regimes from Fig.~\ref{fig:introductory}a and the error bars are $1\sigma$ confidence intervals of the fitted values.}
	\label{FigUnitarity}
\end{figure}

At unitarity the crossover time $t_a$ diverges, such that the universal $t^{3/2}$ dynamics dominates the entire two-body scattering regime. Indeed, the initial amplitude and phase evolution shown in Fig.~\ref{FigUnitarity} agree very well with the dynamics predicted by Eq.~\eqref{eq:Coherence_2body}, confirming both the characteristic decay exponent and the associated time constant $t_n$. This agreement highlights the importance of two-body dynamics over a significant timespan of initial relaxation, even in the unitary limit. 

For longer evolution times,  we observe pronounced deviations from Eq.~\eqref{eq:Coherence_2body} signalling the onset of many-body correlations due to the strong interaction between the impurity state and the condensate. This behavior is captured by the diagrammatic prediction, which yields an excellent description of the non-equilibrium dynamics of impurities in the regime of strong interactions and thus demonstrates the many-body nature of the long-time impurity evolution in our experiments. In particular, the data reveals a clear crossover between the initial two-body $t^{3/2}$ dynamics and a slower many-body decay at a transition time indicated as a white data point in Fig.\ref{fig:introductory}a.

Moreover, the measured phase evolution allows us to track the instantaneous energy of the impurity. Since the phase evolution $\varphi_\text{C}(t) \to - E_\text{p} t / \hbar$ for long times is governed by the polaron energy $E_\text{p}$, we obtain the instantaneous energy from $E(t)=-\hbar d\varphi_C/dt$. As shown in Fig.~\ref{FigUnitarity}b, the observed impurity energy approaches the expected equilibrium polaron energy. Therefore, our measurements directly display the dynamical emergence of the Bose polaron in the regime of strong interactions.


Our experiment covers all relevant time scales of quasiparticle formation and thus opens up new pathways to study non-equilibrium phenomena in strongly interacting quantum many-body systems. The demonstrated technique will enable investigations of bosonic analogues of Anderson's orthogonality catastrophe~\cite{Knap2012} and transport processes~\cite{Sommer2011,Bardon2014} via time-domain measurements. Similar measurements at repulsive impurity interactions will be able to explore the predicted formation of multi-phonon bound states \cite{Shchadilova2016}. Experiments with higher impurity concentrations will permit the investigation of effective polaron interactions~\cite{Camacho-Guardian2018b}. Such mediated interactions are believed to play a vital role for transport properties of condensed matter systems~\cite{Alexandrov2010}. Ultimately, this may enable the observation of strongly bound bosonic bipolarons~\cite{Camacho-Guardian2018} and their formation in a time-resolved manner. Elucidating the dynamics of induced quasiparticle interactions could prove essential, since strong retardation and relaxation effects \cite{Camacho-Guardian2018b,nielsen2019} may render such bipolarons inaccessible to common spectroscopic methods \cite{jorgensen2016,hu2016}.

\section{Acknowledgements}

We thank L. A. Pe\~{n}a Ardila for helpful discussions. This work was supported by the Villum Foundation, the Carlsberg foundation, the Danish council for Independent Research, the Danish National Research Foundation through the Center of Excellence ''CCQ'' (Grant agreement no.: DNRF156) and T.P acknowledges support through a Niels Bohr Professorship.

\section{Author contributions}
M.G.S., T.G.S., N.B.J. and J.J.A. designed and carried out the experiment. M.G.S. performed the data analysis. K.K.N., A.C.-G., T.P., and G.M.B. provided the theoretical predictions. All authors contributed to writing the manuscript.

\section{Data availability}
The data that support the findings of this study are available from the corresponding
author upon reasonable request.

\section{Code availability}
The code that supports the findings of this study are available from the corresponding
author upon reasonable request.

\section{Competing interest}
The authors declare no competing interests.

\section{Methods}
\textbf{Experimental preparation.}
To study the impurity dynamics, a Bose-Einstein condensate of $^{39}$K atoms in the $|F = 1, m_F = -1 \rangle$ state is prepared in an optical dipole potential \cite{wacker2015}. The evaporation is performed near a Feshbach resonance at $\SI{33.6}{G}$ before ramping the magnetic field to a desired valued close to the interstate Feshbach resonance at $\SI{113.8}{G}.$ The cloud temperature is kept constant at $\SI{50}{nK}$ throughout the measurements and the mean geometrical trap frequency is $2\pi\times \SI{65}{Hz}$ ensuring an average condensate density of $n_{\rm B} = 0.7\times 10^{14}$cm$^{-3}$.
\newline

\textbf{Decoherence.} 
Three additional experimental decoherence processes are included
in the theoretical description of the coherence. To account
for processes due to trap inhomogeneity the coherence amplitude and phase are integrated over the cloud density. The effects of
finite lifetime are included by multiplying the theoretical coherence
amplitude with an exponential function based on an
independently measured loss rate $\Gamma_\text{loss}.$ The magnetic
field noise is included similarly by multiplying the theoretical
coherence amplitude with a decay due to shot-to-shot fluctuations
which is measured independently (Supplementary Information).

\clearpage
\onecolumngrid

\section*{\large Supplementary Information for ``Non-equilibrium quantum dynamics and formation of the Bose polaron''}
\section{Ramsey Interferometry}
The customized Ramsey method employed in the experiment maps the impurity coherence to the final number of atoms in the sample. We first analyze the homogeneous case and then perform a local density approximation (LDA) to analyze the experimentally relevant inhomogeneous gas. To generate impurities, a radio frequency (rf) field

\begin{equation}
H_{\rm rf}(\varphi) = \hbar\Omega \sum_{\bk} \left[ \te^{+i\varphi} c_{\bk}^\dagger b_\bk + \te^{-i\varphi} b_\bk^\dagger c_{\bk} \right]
\label{eq.H_rf}
\end{equation}

drives transitions between two magnetic states, transferring atoms from the medium $\ket{b}$ state to the impurity $\ket{c}$ state. The operators $b_{\bk}^\dagger, c^\dagger_{\bk}$ create an atom in the medium and impurity states with momentum $\bk$, respectively, and $H_{\rm rf}$ is given in the rotating frame with the Rabi frequency $\Omega$ and phase $\varphi$. The system Hamiltonian for resonant transfer -- in the rotating frame -- is given by
\begin{align}\label{eq.H_system}
	H =& \sum_{\bk}\varepsilon_\bk \left( c^\dagger_{\bk}c_{\bk} + b^\dagger_\bk b_\bk\right) + \frac{\TB}{2V} \sum_{\bk, \bq, \bp} b^\dagger_{\bk + \bp}b^\dagger_{\bq - \bp} b_{\bq}b_{\bk} + \frac{\T}{V} \sum_{\bk, \bq, \bp} b^\dagger_{\bk + \bp}c^\dagger_{\bq - \bp} c_{\bq}b_{\bk},
\end{align} 
with $\epsilon_\bk = \hbar^2 \bk^2/2m$, system volume $V$, and $\T = 4\pi \hbar^2 a / m, \TB = 4\pi \hbar^2 a_{\rm B} / m$ the zero energy scattering matrices for the impurity-boson and boson-boson interactions respectively. Here we assume that only a single impurity is present, neglecting any impurity-impurity interactions. The Ramsey sequence consists of two short rf-pulses as described in the main text. Since the duration of these pulses is much shorter than the impurity dynamics investigated, we can safely split the time evolution operator into three separate parts $\calU_{\rm tot}(t) = \calU_{\rm rf}(\varphi, \delta t) \calU \calU_{\rm rf}(0, \delta t)$. Here $\calU_{\rm rf}(\varphi, t) = \te^{-iH_{\rm rf}(\varphi)t}$ and $\calU = \te^{-iHt}$. In the first pulse we drive at zero phase, in the second at some variable \textit{probe} phase, $\varphi$. To stay in the single impurity limit we require $\Omega \delta t \ll 1$, which in turn means that we can expand the rf evolution operator
\begin{equation}
\calU_{\rm rf}(\varphi, \delta t) \simeq 1 - i H_{\rm rf}(\varphi) \delta t - \frac{\left(H_{\rm rf}(\varphi)\delta t\right)^2}{2},
\label{eq.U_rf_2nd_order_expansion}
\end{equation}
to second order in $\Omega \delta t$. The initial state of the system is the ground state $\ket{\BEC}$ of $H$ with no impurities present. Using the time evolution operator $\calU_{\rm tot}(t)$ together with the expansion \eqref{eq.U_rf_2nd_order_expansion}, we obtain the mean number of atoms in the impurity state after the two rf-pulses

\begin{align}
	N_{\rm c}(t) &= \bra{\BEC}\calU^\dagger_{\rm tot}(t) \sum_{\bk} c^\dagger_\bk c_\bk \calU_{\rm tot}(t)\ket{\BEC} = N_{\rm B} \cdot 2(\Omega \delta t)^2\Re\left[1 + \te^{-i \varphi} \cdot iG_{\rm bc}(t)\right],
	\label{eq.mean_groundstate_atomnumber}
\end{align}

which is \textit{exact} to second order in $\Omega \delta t$. Here $N_{\rm B}$ is the initial total number of atoms in the $\ket{b}$ state, and 
\begin{equation}
G_{\rm bc}(t) = -\frac{i}{N_{\rm B}}\sum_{\bk, \bq} \bra{\BEC}b^\dagger_\bk(t) c_{\bk}(t) c^\dagger_{\bq}(0)b_{\bq}(0)\ket{\BEC} \nn
\end{equation}
is an impurity-boson Green's function with $c_{\bk}(t) = \calU^\dagger(t) c_{\bk}(0)\calU(t)$ the time evolved annihilation operator for the impurity -- likewise for $b_\bk(t)$. Since the medium atoms are condensed in the zero momentum mode, the dominant contribution to $G_{\rm bc}$ comes from $k = q = 0$. Additional contributions are suppressed by at least a factor of $1 / \sqrt{N_{\rm B}}$. 

We therefore find
\begin{gather}
	G_{\rm bc}(t) \simeq -\frac{i}{N_{\rm B}} \bra{\BEC} b_0^\dagger(t) c_0(t) c_0^\dagger(0) b_0(0)\ket{\BEC} \simeq -i \bra{\BEC} c_0(t) c_0^\dagger(0) \ket{\BEC} = G_0(t), 
\end{gather}
using $b_0\ket{\BEC} \simeq \sqrt{N_{\rm B}} \ket{\BEC}$. We also use $\bra{\BEC} b_0^\dagger(t) \simeq \bra{\BEC} b_0^\dagger(0) \simeq \sqrt{N_{\rm B}} \cdot \bra{\BEC}$. This assumes that the impurity dynamics has little effect on the condensate reservoir. This is well justified for a small fraction of impurities, where corrections are again expected to scale as $1 / \sqrt{N_{\rm B}}$.
Finally, using that $i G_0(t) = C(t) / C(0)$ \cite{nielsen2019}, we obtain a mapping between the impurity density and coherence $C(t)$

\begin{equation}
n_{\rm c}(t) = n_{\rm B} \cdot 2(\Omega \delta t)^2\Re\left[1 + \te^{-i\varphi} \cdot C(t)\right],
\label{eq.mean_number_link_to_coherence}
\end{equation}

by dividing out the system volume, $V$, setting $C(0) = 1$, and defining the initial atom density $n_{\rm B} = N_{\rm B} / V$. 

In the experiment, the atomic gas is held in a harmonic trap $V(\br) = m (\omega_x^2 x^2 + \omega_y^2 y^2 + \omega_z^2 z^2) / 2$. As a result the atomic density is spatially dependent and we adjust the analysis above by using LDA. In a standard Thomas-Fermi approximation this leads to the density $n_{\rm B}(\br) = (\mu - V(\br)) / \TB$, where $\mu$ is the chemical potential of the condensate. In the local density approximation, Eq.~\eqref{eq.mean_number_link_to_coherence} is replaced by the local equation $n_{\rm c}(\br, t) = n_{\rm B}(\br) \cdot 2(\Omega \delta t)^2 \Re [1 + \te^{-i\varphi} \cdot C(\br, t) ]$, where $C(\br, t)$ is the local coherence. The number of impurities after the two rf-pulses is then
\begin{align}
	N_{\rm c}(t) &= \int {\rm d}^3 r \; n_{\rm c}(\br, t) = N_{\rm B} \cdot 2(\Omega \delta t)^2 \Re\left[1 + \te^{i\varphi} \cdot C(t)\right], \nn
\end{align}
defining the trap averaged coherence $C(t) = \int {\rm d}^3 r \; n(\br) C(\br, t) / N_{\rm B}$. Subsequent to the second rf-pulse the atoms are held in the trap, allowing three body recombination to take place, eventually resulting in the loss of two medium atoms for every impurity. The final remaining number of atoms in the system is thus

\begin{align}
	N &= N_{\rm B} - 3 N_{\rm c} = N_{\rm B}\left(1 - 6(\Omega \delta t)^2 \Re\left[1 + \te^{-i\varphi} \cdot C(t)\right]\right) \nn \\
	&= N_0 - 6 N_{\rm B} (\Omega \delta t)^2 |C(t)|\cos(\varphi - \varphi_C(t)),
	\label{eq.mean_number_and_coherence}
\end{align}

with $N_0 = N_{\rm B} (1 - 6(\Omega \delta t)^2)$ the average number of atoms measured as a function of the probe phase $\varphi$ for every evolution time $t$. To enable the experimental analysis the coherence is expressed in terms of its amplitude and phase: $C(t) = |C(t)|\te^{i\varphi_C(t)}$. By performing a fit $N(\varphi) = N_0 - \mathcal{A} \cos(\varphi - \varphi_C)$ to the measured data, we thus extract the phase and the normalized coherence amplitude $|C(t)| = |\mathcal{A}(t) / \mathcal{A}(0)|$ simultaneously. 

\section{Theoretical description of impurity dynamics}
The impurity coherence is in general equal to the Fourier transform of the impurity spectral function $A(\omega)$ at zero momentum
\begin{equation}
C(t) = \int_{-\infty}^{+\infty} \frac{{\rm d}\omega}{2\pi} \, \te^{-i\omega t} A(\omega).
\label{eq.Greens_function_fourier_transform}
\end{equation}
Our approach to predict impurity dynamics is to calculate the spectral function $A(\omega)$ and then determine the dynamics of the coherence. In this section, we present the theoretical description of different regimes of impurity dynamics, from universal short-time behaviour to the non-perturbative treatment of a polaron formation.

\subsection{Initial two-body dynamics}
\label{sub: Universal short-time behaviour}
We start by analysing the short-time behaviour of the coherence. The integral in Eq.~\eqref{eq.Greens_function_fourier_transform} is split as follows
\begin{align}
	C(t) &= \int_{-\infty}^{\infty} \frac{{\rm d}\omega}{2\pi} \, (1 - i\omega t) A(\omega) + \int_{-\infty}^{\infty} \frac{{\rm d}\omega}{2\pi}\, \left(\te^{-i\omega t} - (1 - i\omega t)\right) A(\omega).
	\label{eq.C_short_time_integral_form_0}
\end{align}
It is then apparent that we can use the so-called sum rules \cite{braaten2010} 
\begin{equation}
\int_{-\infty}^{\infty} \frac{{\rm d}\omega}{2\pi} \,  A(\omega) = 1, \hspace{0.5cm} \int_{-\infty}^{\infty} \frac{{\rm d}\omega}{2\pi} \,  \omega A(\omega) = \frac{a_{\rm B}^{-1} - a^{-1}}{4\pi m}\frac{\hbar C_2}{N_{\rm B}},
\label{eq.sum_rules}
\end{equation}
here appropriately rewritten in terms of the spectral function, to calculate the first term in Eq.~\eqref{eq.C_short_time_integral_form_0}. The two-body contact of the BEC, $C_2 =  8\pi m a_{\rm B}^2 / \hbar^2 \cdot {\rm d}E_{\BEC}/{\rm d}\aB = N_{\rm B} \cdot 16 \pi^2 \nB a_{\rm B}^2$, is obtained using Bogoliubov theory appropriate for weak interactions in the condensate, $\nB\aB^3 \ll 1$. The second term in Eq.~\eqref{eq.C_short_time_integral_form_0} can be evaluated at short times, since the factor $\te^{-i\omega t} - (1 - i\omega t)$ removes the low energy sector up to order $(\omega t)^2$. Therefore, at sufficiently short times, one can use the asymptotic behavior of the spectral function at large frequencies calculated in \cite{braaten2010} 
\begin{align}
	\lim_{\omega \to \infty} A(\omega) &= \frac{1}{2\pi}\frac{C_2}{N_{\rm B}} \sqrt{\frac{\hbar}{m}} \frac{\left(a / \aB - 1\right)^2}{1 + m a^2 \, \omega / \hbar} \cdot \frac{1}{\omega^{3/2}}  = \frac{K}{1 + \omega t_a} \cdot \frac{1}{\omega^{3/2}},
	\label{eq.A_asymptotic_behaviour}
\end{align}
with $K = 4 / 3\pi \cdot (1 - \aB/a )^2 (k_n |a|)^3 / \sqrt{t_a}$, $k_n = (6\pi^2 n_{\rm B})^{1/3}$ and $t_a = ma^2 / \hbar$. The $\omega < 0$ part of the second term in Eq.~\eqref{eq.C_short_time_integral_form_0} is negligible for negative impurity-boson scattering lengths \textit{and} close to unitarity. Essentially, the only important contribution in this region is due to the impurity-boson molecular state, which is absent for $a < 0$ and has an energy $E = -\hbar^2 / ma^2$ which goes to zero as we approach unitarity $a \to \infty$. We can thus write
\begin{align}
	C(t) &\simeq \int_{-\infty}^{\infty} \frac{{\rm d}\omega}{2\pi}\, (1 - i\omega t) A(\omega) + \int_{0}^{\infty}\! \frac{{\rm d}\omega}{2\pi} \, \left(\te^{-i\omega t} - (1 - i\omega t)\right) A(\omega) \nn \\
	&\simeq 1 - it \cdot \frac{\nB \TB}{\hbar}\left(1 - \frac{\aB}{a}\right) + K \underset{I}{\underbrace{\int_{0}^{\infty}\ \frac{{\rm d}\omega}{2\pi} \, \frac{\te^{-i\omega t} - (1 - i\omega t)}{1 + \omega t_a} \cdot \frac{1}{\omega^{3/2}} }},
	\label{eq.C_short_time_integral_form}
\end{align}
using $\hbar C_2 / N_{\rm B} \cdot (a_{\rm B}^{-1} - a^{-1}) / 4\pi m = \nB\TB (1 - \aB / a) / \hbar$. The integral $I$ is evaluated using the dimensionless variables $\tilde{\omega} = \omega t_a$ and $\tilde{t} = t / t_a$
\begin{align}
	I &= \frac{\sqrt{t_a}}{2} \left[1 + i \frac{t}{t_a} - \frac{2}{\sqrt{\pi}} \te^{i t / t_a} \Gamma\left(\frac{3}{2}, i\frac{t}{t_a}\right)\right], \nn 
\end{align}
where $\Gamma$ is the incomplete gamma function. Reinserting $I$ into Eq.~\eqref{eq.C_short_time_integral_form} we obtain the short-time behavior of the impurity coherence for general interaction strengths

\begin{align}
	C(t) \simeq & 1 - i \frac{E_{\rm mf}t}{\hbar}  + \frac{2}{3\pi}(k_n|a|)^3\left[1 - \frac{2}{\sqrt{\pi}} \te^{i t / t_a} \Gamma\left(\frac{3}{2}, i\frac{t}{t_a} \right)\right],
	\label{eq.short_time_general}
\end{align}

defining the mean field energy $E_{\rm mf} = \nB \T$ and neglecting $a_{\rm B} / a$ corrections. At very short times, $t \ll t_a = ma^2 / \hbar$, the coherence dynamics has the universal behaviour

\begin{gather}
	C(t)=1 - (1 - i) \frac{16}{9 \pi^{3/2}} \left(\frac{t}{t_n}\right)^{3/2}.
	\label{eq:Coherence_2body1}
\end{gather} 

Since it is independent of the impurity-boson scattering length, $a$, this defines a unitarity limited dynamical regime. 

For weak interactions, the impurity dynamics changes from this two-body unitary dynamics to two-body weak-coupling dynamics governed by

\begin{gather}
	C(t)=1 - iE_{\rm mf}t/ \hbar - (1 + i) \left(\frac{t}{t_{\rm w}}\right)^{1/2}, 
	\label{eq:Coherence_2body2}
\end{gather}
for $\hbar/E_{\text{mf}}\gg t\gg  ma^2/\hbar$, valid to second order in the impurity-boson scattering length. The third-order correction in the impurity-boson scattering length from Eq.~\eqref{eq.short_time_general} leads to the following dynamics
\begin{gather}
	C(t)=1 - iE_{\rm mf}t/ \hbar - (1 + i) \left(\frac{t}{t_{\rm w}}\right)^{1/2}+\frac{2}{3\pi}(k_n|a|)^3, 
	\label{eq:Coherence_2body3}
\end{gather}
where the fourth term becomes relevant for $1/k_n|a|\approx (2/3\pi)^{1/3}\approx 0.59$ which determines the transition from weak to strong interactions.

\subsection{Origin of the two-body power-laws}
The exponent of the power-laws in Eq.~\eqref{eq:Coherence_2body1} and Eq.~\eqref{eq:Coherence_2body2} can be traced back to the cross section $\sigma(k)=4\pi a^2/[1+(ka)^2]$
for $s$-wave scattering between the impurity and a boson from the condensate with relative momentum $k$. At time $t$,  the 
characteristic collision energy giving rise to decoherence is $E\sim\hbar/t$ so that $k\sim \sqrt{m/\hbar t}$ and the typical velocity of the colliding particles is $v\sim\sqrt{\hbar/mt}$. 

At short times $t\ll ma^2/\hbar$ these collisional energies are correspondingly high $E\gg \hbar^2/ma^2$ such that $ka \gg 1$ and the cross section is unitarity-limited $\sigma(k)\simeq 4\pi/k^2\sim\hbar t/m$. Using that the decoherence rate is related to the 
collision rate  $\dot C(t)\sim -n_B\sigma v$, we obtain  $\dot C(t)\sim-n_B(\hbar/m)^{3/2}\sqrt t$ and integrating while setting $C(0)=1$ yields precisely the 
$(t/t_n)^{3/2}$ power-law given in Eq.~\eqref{eq:Coherence_2body1}. On the other hand, for longer times $t\gg ma^2/\hbar$
decoherence involves low energy collisions for which the cross section is $\sigma(k)\simeq 4\pi a^2$. The same line of reasoning then 
gives the $\sqrt{t/t_w}$ power law in Eq.~\eqref{eq:Coherence_2body2}.

\subsection{Many-body dynamics}
To determine the impurity dynamics at arbitrary times, we employ a non-perturbative approach based on the so-called ladder approximation, which includes Feshbach physics via the scattering of one boson out of the condensate by the impurity~\cite{Rath2013}. For $k_na_B\approx 0.01,$ the relevant physics can be explained by assuming an ideal BEC, where the impurity self-energy is $\Sigma(\omega)=n_{\text{B}}{\mathcal T}(\omega)$, with the scattering matrix ${\mathcal T}(\omega)$ in the ladder approximation and the density $n_B$ of the BEC. This yields the spectral function
\begin{align}
	A(\omega)=Z_P2\pi\delta(\omega-\omega_P) + 8\pi  \frac{\hbar^{3/2}n_{\text{B}}}{m^{3/2}\omega^{5/2}} \cdot \frac{\Theta(\omega)}{1+\frac{\hbar}{ma^2\omega}\left(1-\frac{4\pi\hbar n_{\text{B}} a}{m\omega}\right)^2}
	\label{SpectralFnTmatrix}
\end{align}
for zero temperature. Here $\delta(x)$ is the Dirac delta function, $\Theta(\omega)$ is the Heaviside step function, $\hbar \omega_P$ is the polaron energy and $Z_P$ is the polaron residue determined from $\omega_P=\Sigma(\omega_P)$ and $Z_P^{-1}=1-\partial_\omega\Sigma(\omega)|_{\omega_P}$.  Equation~\eqref{SpectralFnTmatrix}	recovers the exact result for large $\omega$ in Eq.~\eqref{eq.A_asymptotic_behaviour}. In addition, it yields a prediction for the low-energy behaviour governed by many-body physics. Specifically, many-body corrections are given by the continuum of high-momentum impurity states and Bogoliubov excitations together with the polaron delta-function peak. The onset of the continuum of states described by the second term in Eq.~\eqref{SpectralFnTmatrix}  starts above the polaron peak instead of at $\omega=0$, since one can make  states with arbitrarily small excitation energy, consisting of a moving polaron and a Bogoliubov mode with a total momentum of zero. In our predictions, we therefore employ a step in a self-consistent calculation which moves the continuum to start just above the polaron peak. In addition, we add a small imaginary width $i\eta=0.05E_n$. Moreover, the theory is averaged over the trap to model the experiment. Finally, by Fourier transforming the result we obtain the prediction for the coherence which is shown in Fig.~2, 3 and 4 of the main manuscript.

\begin{figure}[h]
	\begin{center}
		\includegraphics[width=10cm]{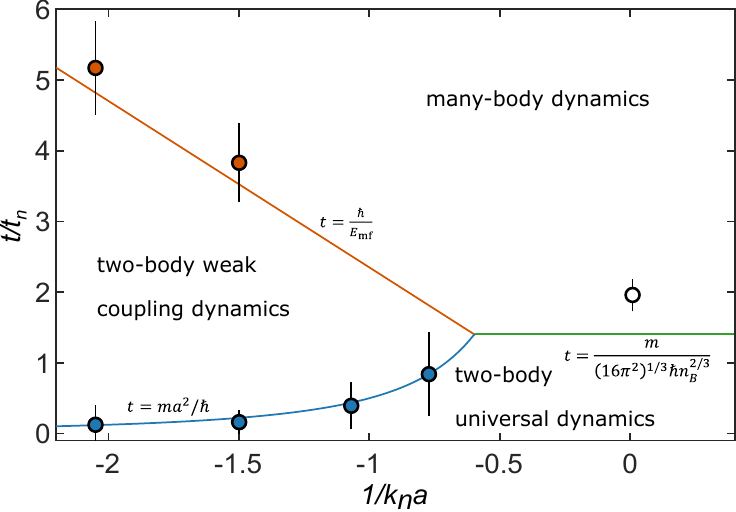}
	\end{center}
	\caption{Regimes of impurity dynamics as a function of inverse interaction strength $1/k_n a$ and evolution time $t/t_n$. The lines show the transition times separating the different dynamical regimes: The universal to weak-coupling transition is shown in blue, the  weak-coupling to many-body transition is shown in orange, and the universal to many-body transition is shown in green. The points show the experimentally observed transition times and the error bars correspond to the $1\sigma$ confidence intervals of the fitted values (blue) and the data resolution (red and white) as described in Sec.~\ref{sub: exp transition times}. We emphasize that although the transitions are shown as lines, they are in reality smooth as indicated in Fig.~1a in the main manuscript.}
	\label{FigDiag} 
\end{figure} 

\subsection{Dynamical regimes of impurity evolution}
We can now identify various regimes of the impurity dynamics by comparing  the relative magnitude of the terms in the denominator of Eq.~\eqref{SpectralFnTmatrix}. The result is shown in Fig.~\ref{FigDiag} where the transitions between different regimes are smooth and should not be understood as sharp boundaries. For high energies, corresponding to short times, the spectral function scales as $\omega^{-5/2}$ giving the universal $t^{3/2}$ dynamics described by Eq.~\eqref{eq:Coherence_2body1}. For weak interactions $(E_{\text{mf}}\ll  \hbar^2/ma^2)$, there is a transition to the $t^{1/2}$ dynamics also described by Eq.~\eqref{eq:Coherence_2body2} for $t\gtrsim t_a = ma^2/\hbar$ and eventually many-body physics sets in at $t \gtrsim  \hbar / E_{\rm mf}$. 
Since $ma^2/\hbar$ establishes the crossover between the different regimes of two-body impurity dynamics, the transition to Eq.~\eqref{eq:Coherence_2body2} is restricted to values of the interaction strengths such that  $ma^2/\hbar$ remains the shortest time-scale of the system. Using the condition $t_a=\hbar/E_{\rm mf}$ this holds true for weak and intermediate interaction strength until $1/(k_n|a|)=(2/3\pi)^{1/3},$ in agreement with Eq.~\eqref{eq:Coherence_2body3}.   
For strong interactions  $(E_{\text{mf}}\ge  \hbar^2/ma^2)$, this is no longer the case and the crossover to the  $t^{1/2}$ dynamics in Eq.~\eqref{eq:Coherence_2body2} is prevented. Here, the coherence transitions directly from the universal $t^{3/2}$ dynamics to the many-body dynamics at $t \gtrsim (16\pi^2)^{-1/3}mn_B^{-2/3}/\hbar = (3\pi / 2)^{2/3} / 2 \cdot t_n \simeq 1.4 \;t_n$, as shown in Fig.~\ref{FigDiag}.

\section{Decoherence effects}
\label{Sec_Decoherence}

To accurately describe the evolution of the coherence, every prominent technical source of decoherence must be considered. This section accounts for three decoherence mechanisms which are included in our theoretical description of the data.

\begin{figure}[tb]
	\centering
	\includegraphics[width=16cm]{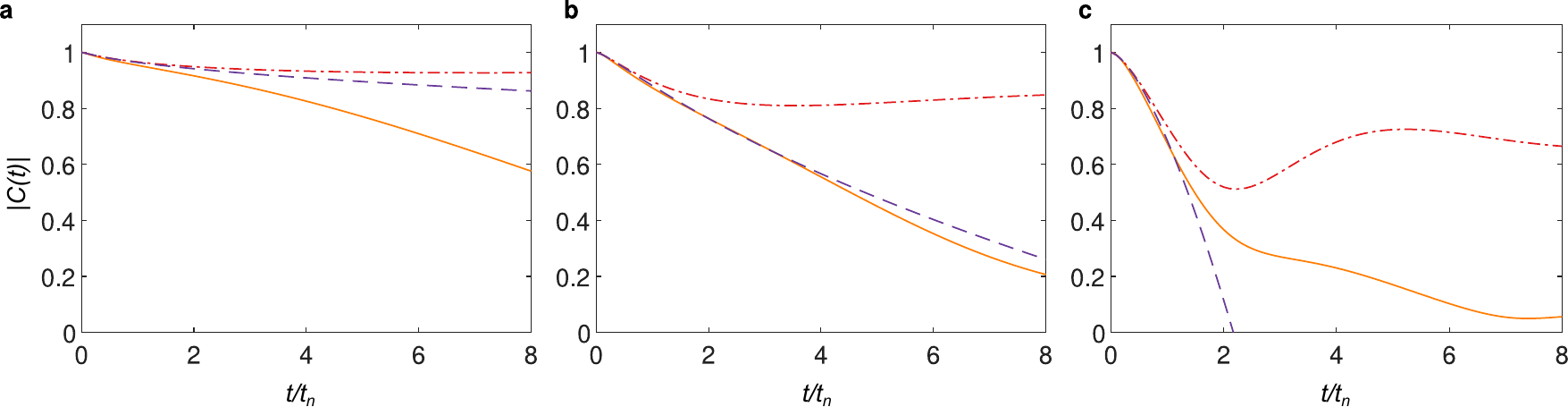}
	\caption{Loss of coherence due to the harmonic trap potential. The coherence amplitude for (a) $1/k_n a = -2$, (b) $1/k_n a = -0.77$ and (c) $1/k_n a = 0.01$ are shown for the general two-body description Eq.~\eqref{eq.short_time_general} as a dashed purple line and the diagrammatic prediction with and without trap dephasing as a solid orange and dash-dotted red line, respectively. No other technical decoherence sources are included for this comparison.}
	\label{fig_sm:homogeneous}
\end{figure}

\subsection{Decoherence from harmonic trap}

The harmonic potential provided by the optical dipole trap results in an inhomogeneous density distribution of the atoms. Since the impurity population is created evenly across the condensate, the density dependent interaction strength results in dephasing of the system. We therefore have to integrate over the spatially dependent terms of the coherence weighted by the density distribution. For the short-time two-body theoretical predictions in Sec.~\ref{sub: Universal short-time behaviour}, this simply corresponds to replacing the density distribution with its average value, since all terms are linear in density. For the dynamics occurring at longer times, the density dependence becomes non-linear and the averaging induces a more subtle decoherence process of the impurity dynamics.  For weak interactions, the trap-averaged dynamics and the homogeneous dynamics both signal the onset of many-body physics as they deviate from the two-body prediction at similar times, as shown in Fig.~\ref{fig_sm:homogeneous}a. For intermediate interaction strengths, dephasing suppresses the trap-averaged prediction and it deviates faster from the homogeneous dynamics than from the two-body theory. Subsequently, the trap-averaged many-body and the two-body theory agree for longer times than the transition time predicted by the homogeneous dynamics, as illustrated in Fig.~\ref{fig_sm:homogeneous}b. For strong interactions, the dephasing only affects impurity dynamics at times larger than the transition time from universal to many-body dynamics, and therefore, it is possible to extract the transition from the trap-averaged data, as shown in Fig.~\ref{fig_sm:homogeneous}c.

\subsection{Decoherence from finite impurity lifetime}

Strongly interacting Bose gases are typically subject to rapid loss from inelastic three-body decay. In Ramsey interferometry, such decay processes result in a loss of contrast and must therefore be taken into account. In our experimental system the impurity lifetime is typically shorter than the time of flight and all impurity population is lost before it can be observed using absorption imaging after expansion. 

We therefore employ a more sophisticated strategy to measure the impurity loss rate: A BEC is prepared in the  $\ket{F = 1, m_F = -1}$ state under conditions similar to those presented in the main text. To initialize a loss measurement, a rf-pulse transfers approximately $10 \%$  of the population to the impurity state $\ket{F = 1, m_F = 0}$. The sample is then held for a variable time during which three-body recombination processes take place. Subsequently, any remaining population in the impurity state is transferred to the $\ket{F = 1, m_F = 1}$ state by a $\pi$-pulse. The population in this state undergoes two-body spin-changing collisions with the population in the $\ket{F = 1, m_F = -1}$ state and is rapidly lost. Thus, the transferred fraction is always lost, but through different processes depending on the state in which the loss takes place. Finally, the remaining number of BEC atoms is recorded by absorption imaging after expansion.

\begin{figure}[tb]
	\centering
	\includegraphics[width=10cm]{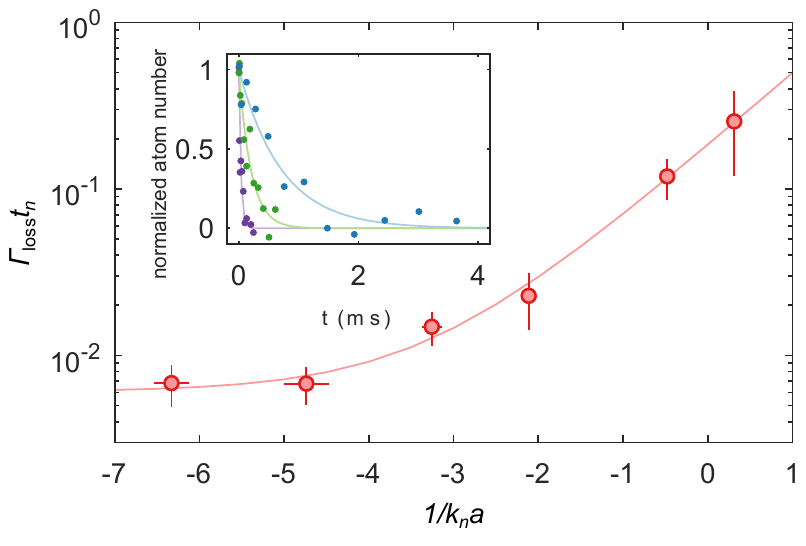}
	\caption{Loss rate of the impurity state as a function of inverse interaction strength. The inset shows selected data and fits for $1/k_na = -4.7$ (blue), $-2.1$ (green), and $-0.48$ (purple). The data have been scaled with the initial and final atom numbers. The main panel shows the obtained loss rates and an empirical fit. The horizontal error bars are due to statistical uncertainty of the atom number and fitting errors from trap frequency measurements. The vertical error bars additionally contain propagated fitting errors from impurity lifetime measurements.
	}
	\label{fig_sm:lifetime}
\end{figure}

Examples of the recorded normalized atom number are shown in the inset of Fig.~\ref{fig_sm:lifetime} for selected interaction strengths $1/k_na$. For each interaction, we perform an exponential fit $\sim \exp (- \Gamma_\text{loss} t)$, with $\Gamma_\text{loss}$ being the loss rate of the impurity state. The obtained loss rates are shown in Fig.~\ref{fig_sm:lifetime}, and as expected, the loss rate increases with the interaction strength. 

These observations are relevant in relation to recent Bose polaron observations~\cite{jorgensen2016, hu2016}. Importantly, the observed loss rate $\hbar\Gamma_\text{loss}$ is smaller than the Bose polaron energies observed in the same system~\cite{jorgensen2016, ardila2019}. At unitarity, the loss rate is comparable to loss rates observed for Bose polarons in a $^{40}$K$^{87}$Rb mixture. However, at intermediate interactions, it is interesting to note that the rate is significantly larger in the $^{39}$K system. This difference in loss rates in the two different atomic systems is also found when comparing three-body loss rates of thermal KRb mixtures~\cite{bloom2013, wacker2016} to single-component thermal $^{39}$K~\cite{zaccanti2009, roy2013, wacker2018}. We therefore conclude that this difference is primarily a consequence of the three-body loss rates of the individual atomic systems.

To model the influence of the observed loss rate on the impurity coherence, we perform an empirical fit $\beta_1 + \beta_2\exp(\beta_3/k_na)$, with fitting parameters $\beta_i$, which is shown in Fig.~\ref{fig_sm:lifetime}. The fit follows the experimental data well, and we therefore employ this function to calculate $\Gamma_\text{loss}$ for arbitrary interactions. To compare with experimental results this loss is included in the theoretically calculated coherence as $C(t) \rightarrow C(t)\exp (- \Gamma_\text{loss} t)$.

\subsection{Decoherence from magnetic field fluctuations}

Experimentally shot-to-shot fluctuations of the magnetic field lead to a further decoherence mechanism. The effect only provides significant decoherence at long times compared to $t_n$ and is therefore mainly relevant for data acquired at weak interactions.

The central part of the experimental procedure is the Ramsey interferometry sequence which is repeated multiple times for each set of experimental parameters. For each repetition, however, the interferometry pulse has a different detuning $\Delta$ compared to the bare transition, due to shot-to-shot fluctuations of the magnetic field. This detuning thus provides an additional phase shift $2 \pi \Delta \cdot t$, and when the Ramsey interferometry sequence is repeated several times, these varying phase shifts lead to additional decay of the coherence function. 

\begin{figure}[tb]
	\centering
	\includegraphics[width=10cm]{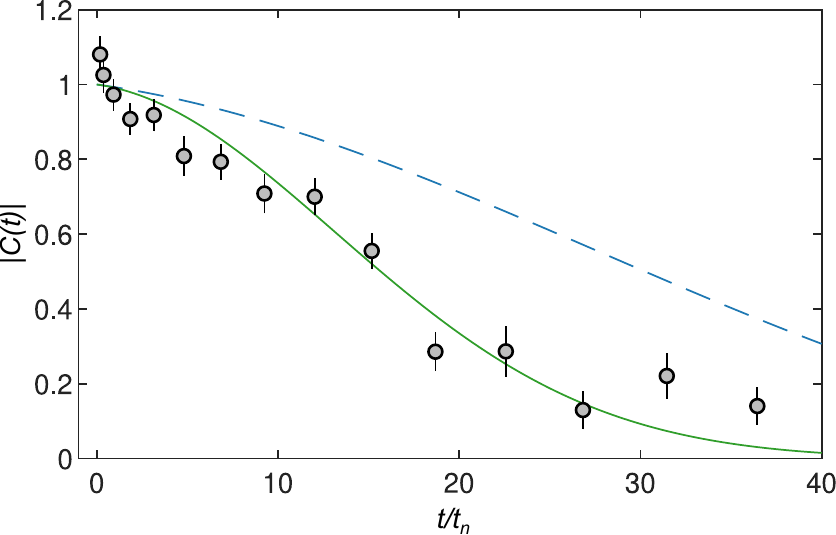}
	\caption{Coherence amplitude at inverse interaction strength $1/k_n a = -5$. The dashed blue line shows the expected amplitude due to the inhomogeneous density distribution and the effect of finite impurity lifetime. The green line is a fit including shot-to-shot fluctuations in the magnetic field yielding $\Delta_\text{noise} = \SI{1.8(1)}{kHz}$. The error bars are $1\sigma$ confidence intervals of the fitted values.
	}
	\label{fig_sm:BfieldNoise}
\end{figure}

To quantify this effect, we assume that $\Delta$ follows a normal distribution, which results in a phase distribution given by $1/\sqrt{2\pi \sigma_\text{noise}^2 (t)} \exp [ - \phi^2/2\sigma_\text{noise}^2 (t)]$, where $\sigma_\text{noise} (t) = 2 \pi \Delta_\text{noise} t$ and $\phi$ is the additional phase. The effect of magnetic field fluctuations on the coherence is then obtained by integrating the phase distribution
\begin{align}
	C(t) \rightarrow \frac{C(t)}{\sqrt{2\pi \sigma_\text{noise}^2 (t)}} \int_{-\infty}^{\infty} \exp(-i\phi) \exp [ - \phi^2/2\sigma_\text{noise}^2 (t)] d\phi.
	\label{eq:bfield_dec}
\end{align}
To obtain the magnitude of $\Delta_\text{noise}$, we have performed Ramsey interferometry measurements at weak interactions $1/k_na = -5$. Here, decoherence from higher-order impurity dynamics is negligible, and the loss of coherence is thus determined by the inhomogeneous density distribution in the trap, finite impurity lifetime, and decoherence due to magnetic field fluctuations. The observed coherence amplitude is shown in Fig.~\ref{fig_sm:BfieldNoise}. We perform a fit according to Eq.~\eqref{eq:bfield_dec} with $\Delta_\text{noise}$ as a fitting parameter and obtain $\Delta_\text{noise} = \SI{1.8(1)}{kHz}$. This effects is included in the theoretical results at all interaction strengths.

\section{Impurity fraction}
The experiment is based on interferometric measurements using a Ramsey-type sequence. To obtain a low population of the impurity state we retain the Bloch vector close to the north pole. The impurity fraction is calibrated through initial Rabi measurements using thermal atoms. A $\pi/7$ pulse is chosen, corresponding to $\sim 5\%$ population in the impurity state. Even though this fraction is low the finite amount of impurities may give rise to interactions between them. It is therefore important to examine if such impurity-impurity interactions influence the experiment.

To investigate this, additional measurements of the coherence were performed at $1/k_na = -1$ for $5\%$, $15\%$ and $20\%$ impurity fraction of the total atom number. However, no significant change in coherence amplitude or phase was observed and therefore a fraction of $5\%$ was chosen for the experiments presented in the main manuscript. A similar investigation of the impurity fraction for the spectral response of the Bose polaron was previously performed in this system~\cite{jorgensen2016}. In that case, no significant effect was observed up to $25\%$ impurity fraction and $10\%$ were chosen for those spectroscopic measurements.

\section{Experimental data analysis}
In this section the main elements of the data analysis are presented. They consist of the normalization of the coherence amplitude, a discussion of the coherence amplitude and phase evolution, the extraction of the boundaries between the dynamical regimes of impurity dynamics, and the calculation of the instantaneous energy.
\subsection{Experimental normalization of coherence amplitude}

At weak interactions, we observe that the atom number loss from the BEC is consistent with three-body recombination between one impurity atom and two medium atoms. However, towards stronger interactions, we observe an increased loss, which is likely due to higher-order losses under these conditions. This hinders a simple conversion between the amplitude of the BEC atom number oscillations and the coherence amplitude, since a new proportionality factor is required at each interaction strength. 

Instead, we employ the general short-time model of Eq.~\eqref{eq.short_time_general}. For each data set, we fit the measured coherence amplitude with Eq.~\eqref{eq.short_time_general} within $\sim \SI{10}{\micro s}$ and obtain the initial amplitude $\mathcal{A}(0)$, which is used to scale the measured coherence amplitude. Note that this normalization procedure does not influence the coherence phase $\varphi_C$ or relative amplitudes $|C(t') / C(t)|$.

\subsection{Coherence amplitude and phase evolution}
The three data sets presented in the main manuscript exhibit vastly different dynamical behavior. Here, we elaborate these differences and present the data side by side for a more direct comparison.

\begin{figure}[tb]
	\centering
	\includegraphics[width=16cm]{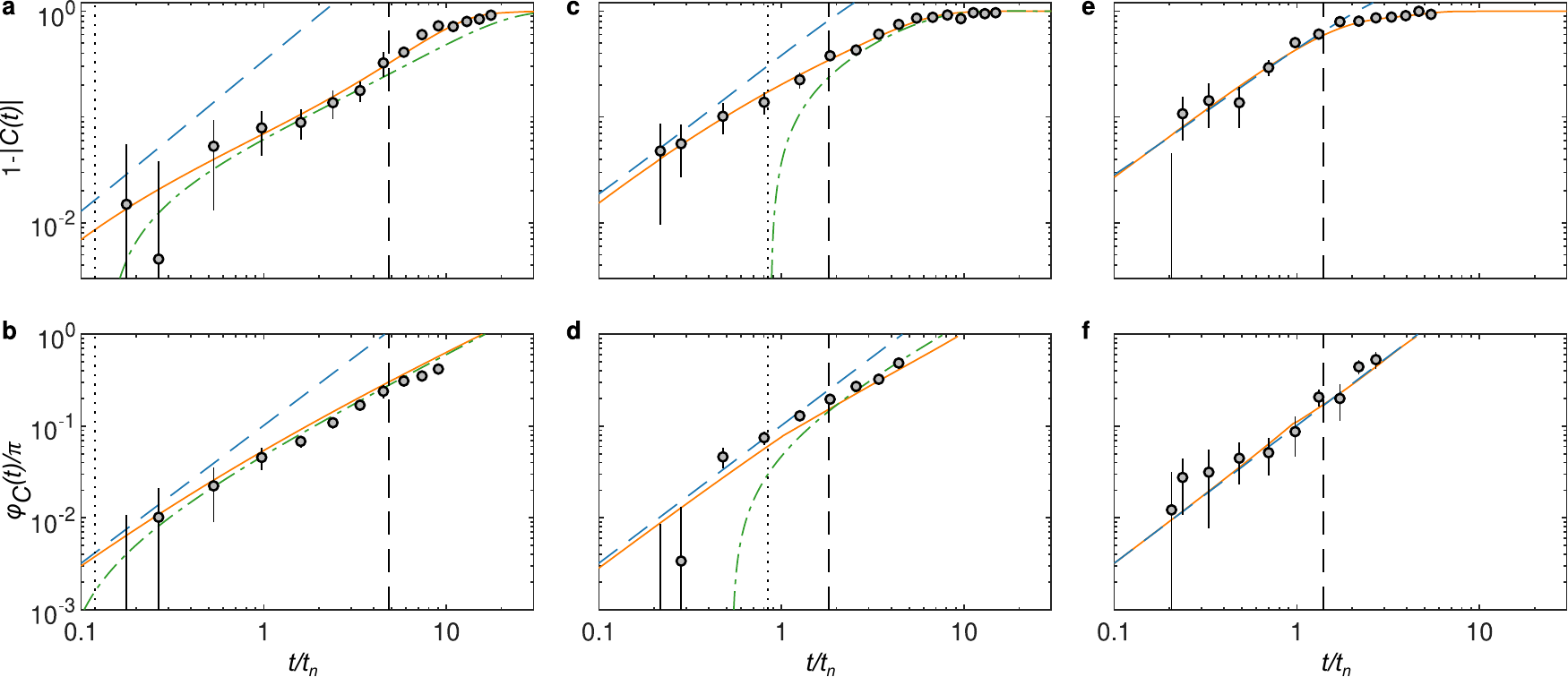}
	\caption{Coherence amplitude and phase evolution at various interaction strengths. Dynamical evolution at (a-b) $1/k_n a = -2$, (c-d) $1/k_n a = -0.77$, and (e-f) $1/k_n a = 0.01$ . The theoretical prediction for two-body universal and two-body weak coupling dynamics are shown as a dashed blue and dash-dotted green line, respectively, and the diagrammatic description is given as a solid orange line. A dotted black line indicates the transition between the two-body regimes and a dashed black line indicates the transition to the many-body regime. The error bars are $1\sigma$ confidence intervals of the fitted values.
	}
	\label{fig_sm:coherence}
\end{figure}

In Fig.~\ref{fig_sm:coherence} the coherence amplitude (top row) and phase evolution (bottom row) for the three interaction strengths discussed in the main manuscript are shown on logarithmic axes. For weak interactions (a-b) the transition from two-body universal to weak coupling dynamics occurs at very short times causing the data to exhibit weak coupling dynamics for a long interval of evolution times until the transition to many-body dynamics. Beyond this transition the coherence continues agreeing with the diagrammatic description, exhibiting many-body behavior which is most pronounced for the coherence amplitude. At intermediate interactions (c-d) the data initially displays clear signs of universal dynamics. Subsequently, the evolution slows down and the data connects with the weak coupling dynamics. However, this  behavior coincides with the transition to the many-body regime, indicating that the smooth transitions between regimes cannot be distinguished when they occur at similar evolution times. Finally, the data at unitarity (e-f) shows clear agreement with the universal prediction before the transition to the regime of many-body dynamics. Subsequently, it continues to follow the diagrammatic description, showing clear many-body behavior which is again most visible for the coherence amplitude.

Generally, the amplitudes can be extracted more reliably even at small signal-to-noise ratio when the phase determination fails. 

\subsection{Experimental determination of the dynamical transitions}
\label{sub: exp transition times}

In Fig.~1 (main manuscript) three distinct regions of dynamical impurity behavior are identified and the boundaries between these regions are shown to agree with the experiment. Here we describe how the displayed data points are obtained experimentally.

The two-body universal and weak-coupling regions can be described by the general two-body short-time equation~\eqref{eq.short_time_general}, while the third region is dominated by many-body physics. For the data at $|1/k_n a| \geq 1.5$ the general two-body expression Eq.~\eqref{eq.short_time_general} fails to agree with the data for times $t \geq 4 t_n$, which indicates that many-body physics starts to dominate the evolution of the coherence. This motivates the following criterion, which is applied to the data sets with $|1/k_n a| \geq 1.5$: The data point at the shortest time which is more than 2 standard errors away from the result of Eq.~\eqref{eq.short_time_general} is identified. The onset of many-body physics then corresponds to the time between this and the previous data point as shown in Fig.~\ref{FigDiag} (orange points). Due to this procedure, the onset and its uncertainty is limited by the experimental resolution.

At unitarity, we investigate when the general two-body expression fails to reproduce the data in a similar manner. We apply the same criterion to the data shown in Fig.~4 of the main manuscript and obtain the data point shown in Fig.~\ref{FigDiag} (white point). 

Finally, we analyse the crossover from two-body universal to two-body weak coupling behavior for all data sets with $|1/k_n a| > 0.5$. To this end, Eq.~\eqref{eq.short_time_general} is fitted to the measured coherence amplitude and phase evolution for the initial dynamics with the transition time $\tilde{t}_a$ as the only free parameter. The fitted values and their errors are also shown in Fig.~\ref{FigDiag} (blue points). Note that some of the data sets do not include data points below the extracted value $\tilde{t}_a$. The term dependent on $t_a$ in Eq.~\eqref{eq.short_time_general}, however, modifies the shape of the curve for times far beyond the time $t_a$ itself, allowing an extraction of this time.

\subsection{Instantaneous energy}
Based on the measured phase of the coherence function, the instantaneous energy of the impurity can be calculated as $E(t)=-\hbar d\varphi_C/dt$. In the mean-field regime, the system equilibrates fast, causing the observed phase evolution to be linear, thus reproducing the constant mean-field energy. For strong interactions, however, the equilibration of the system can be resolved while the impurity state evolves dynamically. The slope of the phase evolution is extracted by piecewise linear fitting to the data in overlapping bins of 4 points, which yields the instantaneous energy and its error. This is shown in Fig.~4b (main manuscript) for the data set obtained at unitarity in good agreement with the time derivative of the ladder approximation theory. Furthermore, the expected polaron energy is plotted based on previously reported experimental results~\cite{ardila2019}.

\bibliographystyle{naturemag}
\bibliography{ImpurityDynamicsBib}

\begin{thebibliography}{10}
\expandafter\ifx\csname url\endcsname\relax
  \def\url#1{\texttt{#1}}\fi
\expandafter\ifx\csname urlprefix\endcsname\relax\def\urlprefix{URL }\fi
\providecommand{\bibinfo}[2]{#2}
\providecommand{\eprint}[2][]{\url{#2}}

\bibitem{Landau1933}
\bibinfo{author}{Landau, L.~D.}
\newblock \bibinfo{title}{{\"Uber die Bewegung der Elektronen in
  Kristalgitter}}.
\newblock \emph{\bibinfo{journal}{Phys. Z. Sowjetunion}}
  \textbf{\bibinfo{volume}{3}}, \bibinfo{pages}{644} (\bibinfo{year}{1933}).

\bibitem{Pekar1946}
\bibinfo{author}{Pekar, S.~I.}
\newblock \bibinfo{title}{Autolocalization of the electron in an inertially
  polarizable dielectric medium}.
\newblock \emph{\bibinfo{journal}{Zh. Eksp. Teor. Fiz.}}
  \textbf{\bibinfo{volume}{16}}, \bibinfo{pages}{335} (\bibinfo{year}{1946}).

\bibitem{Shashi2014}
\bibinfo{author}{Shashi, A.}, \bibinfo{author}{Grusdt, F.},
  \bibinfo{author}{Abanin, D.~A.} \& \bibinfo{author}{Demler, E.}
\newblock \bibinfo{title}{Radio-frequency spectroscopy of polarons in ultracold
  {Bose} gases}.
\newblock \emph{\bibinfo{journal}{Phys. Rev. A}} \textbf{\bibinfo{volume}{89}},
  \bibinfo{pages}{053617} (\bibinfo{year}{2014}).

\bibitem{Shchadilova2016}
\bibinfo{author}{Shchadilova, Y.~E.}, \bibinfo{author}{Schmidt, R.},
  \bibinfo{author}{Grusdt, F.} \& \bibinfo{author}{Demler, E.}
\newblock \bibinfo{title}{Quantum dynamics of ultracold {Bose} polarons}.
\newblock \emph{\bibinfo{journal}{Phys. Rev. Lett.}}
  \textbf{\bibinfo{volume}{117}}, \bibinfo{pages}{113002}
  (\bibinfo{year}{2016}).

\bibitem{nielsen2019}
\bibinfo{author}{Nielsen, K.~K.}, \bibinfo{author}{Ardila, L. A.~P.},
  \bibinfo{author}{Bruun, G.~M.} \& \bibinfo{author}{Pohl, T.}
\newblock \bibinfo{title}{Critical slowdown of non-equilibrium polaron
  dynamics}.
\newblock \emph{\bibinfo{journal}{New Journal of Physics}}
  \textbf{\bibinfo{volume}{21}}, \bibinfo{pages}{043014}
  (\bibinfo{year}{2019}).

\bibitem{Mistakidis2019}
\bibinfo{author}{Mistakidis, S.~I.}, \bibinfo{author}{Katsimiga, G.~C.},
  \bibinfo{author}{Koutentakis, G.~M.}, \bibinfo{author}{Busch, T.} \&
  \bibinfo{author}{Schmelcher, P.}
\newblock \bibinfo{title}{Quench dynamics and orthogonality catastrophe of
  {Bose} polarons}.
\newblock \emph{\bibinfo{journal}{Phys. Rev. Lett.}}
  \textbf{\bibinfo{volume}{122}}, \bibinfo{pages}{183001}
  (\bibinfo{year}{2019}).

\bibitem{Drescher2020}
\bibinfo{author}{Drescher, M.}, \bibinfo{author}{Salmhofer, M.} \&
  \bibinfo{author}{Enss, T.}
\newblock \bibinfo{title}{Theory of a resonantly interacting impurity in a
  {Bose-Einstein} condensate}.
\newblock \emph{\bibinfo{journal}{Phys. Rev. Research}}
  \textbf{\bibinfo{volume}{2}}, \bibinfo{pages}{032011} (\bibinfo{year}{2020}).

\bibitem{Mannella2005}
\bibinfo{author}{Mannella, N.} \emph{et~al.}
\newblock \bibinfo{title}{Nodal quasiparticle in pseudogapped colossal
  magnetoresistive manganites}.
\newblock \emph{\bibinfo{journal}{Nature}} \textbf{\bibinfo{volume}{438}},
  \bibinfo{pages}{474--478} (\bibinfo{year}{2005}).

\bibitem{Lee2006}
\bibinfo{author}{Lee, P.~A.}, \bibinfo{author}{Nagaosa, N.} \&
  \bibinfo{author}{Wen, X.-G.}
\newblock \bibinfo{title}{Doping a mott insulator: Physics of high-temperature
  superconductivity}.
\newblock \emph{\bibinfo{journal}{Rev. Mod. Phys.}}
  \textbf{\bibinfo{volume}{78}}, \bibinfo{pages}{17--85}
  (\bibinfo{year}{2006}).

\bibitem{Bloch2012}
\bibinfo{author}{Bloch, I.}, \bibinfo{author}{Dalibard, J.} \&
  \bibinfo{author}{Nascimb{\`{e}}ne, S.}
\newblock \bibinfo{title}{{Quantum simulations with ultracold quantum gases}}.
\newblock \emph{\bibinfo{journal}{Nature Physics}}
  \textbf{\bibinfo{volume}{8}}, \bibinfo{pages}{267--276}
  (\bibinfo{year}{2012}).

\bibitem{Schirotzek2009}
\bibinfo{author}{Schirotzek, A.}, \bibinfo{author}{Wu, C.-H.},
  \bibinfo{author}{Sommer, A.} \& \bibinfo{author}{Zwierlein, M.~W.}
\newblock \bibinfo{title}{{Observation of Fermi polarons in a tunable Fermi
  liquid of ultracold atoms}}.
\newblock \emph{\bibinfo{journal}{Phys. Rev. Lett.}}
  \textbf{\bibinfo{volume}{102}}, \bibinfo{pages}{230402}
  (\bibinfo{year}{2009}).

\bibitem{Kohstall2012}
\bibinfo{author}{{Kohstall}, C.} \emph{et~al.}
\newblock \bibinfo{title}{{Metastability and coherence of repulsive polarons in
  a strongly interacting Fermi mixture}}.
\newblock \emph{\bibinfo{journal}{Nature}} \textbf{\bibinfo{volume}{485}},
  \bibinfo{pages}{615--618} (\bibinfo{year}{2012}).

\bibitem{Koschorreck2012}
\bibinfo{author}{{Koschorreck}, M.} \emph{et~al.}
\newblock \bibinfo{title}{{Attractive and repulsive Fermi polarons in two
  dimensions}}.
\newblock \emph{\bibinfo{journal}{\nat}} \textbf{\bibinfo{volume}{485}},
  \bibinfo{pages}{619--622} (\bibinfo{year}{2012}).

\bibitem{Massignan2014}
\bibinfo{author}{Massignan, P.}, \bibinfo{author}{Zaccanti, M.} \&
  \bibinfo{author}{Bruun, G.~M.}
\newblock \bibinfo{title}{Polarons, dressed molecules and itinerant
  ferromagnetism in ultracold fermi gases}.
\newblock \emph{\bibinfo{journal}{Reports on Progress in Physics}}
  \textbf{\bibinfo{volume}{77}}, \bibinfo{pages}{034401}
  (\bibinfo{year}{2014}).

\bibitem{cetina2016}
\bibinfo{author}{Cetina, M.} \emph{et~al.}
\newblock \bibinfo{title}{Ultrafast many-body interferometry of impurities
  coupled to a {Fermi} sea}.
\newblock \emph{\bibinfo{journal}{Science}} \textbf{\bibinfo{volume}{354}},
  \bibinfo{pages}{96--99} (\bibinfo{year}{2016}).

\bibitem{Scazza2017}
\bibinfo{author}{Scazza, F.} \emph{et~al.}
\newblock \bibinfo{title}{Repulsive {Fermi} polarons in a resonant mixture of
  ultracold $^{6}\mathrm{Li}$ atoms}.
\newblock \emph{\bibinfo{journal}{Phys. Rev. Lett.}}
  \textbf{\bibinfo{volume}{118}}, \bibinfo{pages}{083602}
  (\bibinfo{year}{2017}).

\bibitem{Schmidt2018}
\bibinfo{author}{Schmidt, R.} \emph{et~al.}
\newblock \bibinfo{title}{Universal many-body response of heavy impurities
  coupled to a {Fermi} sea: a review of recent progress}.
\newblock \emph{\bibinfo{journal}{Reports on Progress in Physics}}
  \textbf{\bibinfo{volume}{81}}, \bibinfo{pages}{024401}
  (\bibinfo{year}{2018}).

\bibitem{Yan2019}
\bibinfo{author}{Yan, Z.} \emph{et~al.}
\newblock \bibinfo{title}{Boiling a unitary {Fermi} liquid}.
\newblock \emph{\bibinfo{journal}{Phys. Rev. Lett.}}
  \textbf{\bibinfo{volume}{122}}, \bibinfo{pages}{093401}
  (\bibinfo{year}{2019}).

\bibitem{Darkwah2019}
\bibinfo{author}{Darkwah~Oppong, N.} \emph{et~al.}
\newblock \bibinfo{title}{Observation of coherent multiorbital polarons in a
  two-dimensional {Fermi} gas}.
\newblock \emph{\bibinfo{journal}{Phys. Rev. Lett.}}
  \textbf{\bibinfo{volume}{122}}, \bibinfo{pages}{193604}
  (\bibinfo{year}{2019}).

\bibitem{jorgensen2016}
\bibinfo{author}{J\o{}rgensen, N.~B.} \emph{et~al.}
\newblock \bibinfo{title}{Observation of attractive and repulsive polarons in a
  {Bose-Einstein} condensate}.
\newblock \emph{\bibinfo{journal}{Phys. Rev. Lett.}}
  \textbf{\bibinfo{volume}{117}}, \bibinfo{pages}{055302}
  (\bibinfo{year}{2016}).

\bibitem{hu2016}
\bibinfo{author}{Hu, M.-G.} \emph{et~al.}
\newblock \bibinfo{title}{Bose polarons in the strongly interacting regime}.
\newblock \emph{\bibinfo{journal}{Phys. Rev. Lett.}}
  \textbf{\bibinfo{volume}{117}}, \bibinfo{pages}{055301}
  (\bibinfo{year}{2016}).

\bibitem{Yan2020}
\bibinfo{author}{Yan, Z.~Z.}, \bibinfo{author}{Ni, Y.},
  \bibinfo{author}{Robens, C.} \& \bibinfo{author}{Zwierlein, M.~W.}
\newblock \bibinfo{title}{Bose polarons near quantum criticality}.
\newblock \emph{\bibinfo{journal}{Science}} \textbf{\bibinfo{volume}{368}},
  \bibinfo{pages}{190--194} (\bibinfo{year}{2020}).

\bibitem{ardila2019}
\bibinfo{author}{Pe\~na Ardila, L.~A.} \emph{et~al.}
\newblock \bibinfo{title}{Analyzing a {Bose} polaron across resonant
  interactions}.
\newblock \emph{\bibinfo{journal}{Phys. Rev. A}} \textbf{\bibinfo{volume}{99}},
  \bibinfo{pages}{063607} (\bibinfo{year}{2019}).

\bibitem{Parish2016}
\bibinfo{author}{Parish, M.~M.} \& \bibinfo{author}{Levinsen, J.}
\newblock \bibinfo{title}{Quantum dynamics of impurities coupled to a {Fermi}
  sea}.
\newblock \emph{\bibinfo{journal}{Phys. Rev. B}} \textbf{\bibinfo{volume}{94}},
  \bibinfo{pages}{184303} (\bibinfo{year}{2016}).

\bibitem{wacker2015}
\bibinfo{author}{Wacker, L.} \emph{et~al.}
\newblock \bibinfo{title}{Tunable dual-species {Bose-Einstein} condensates of
  $^{39}\mathrm{K}$ and $^{87}\mathrm{Rb}$}.
\newblock \emph{\bibinfo{journal}{Phys. Rev. A}} \textbf{\bibinfo{volume}{92}},
  \bibinfo{pages}{053602} (\bibinfo{year}{2015}).

\bibitem{lysebo2010}
\bibinfo{author}{Lysebo, M.} \& \bibinfo{author}{Veseth, L.}
\newblock \bibinfo{title}{Feshbach resonances and transition rates for cold
  homonuclear collisions between $^{39}\mathrm{K}$ and $^{41}\mathrm{K}$
  atoms}.
\newblock \emph{\bibinfo{journal}{Phys. Rev. A}} \textbf{\bibinfo{volume}{81}},
  \bibinfo{pages}{032702} (\bibinfo{year}{2010}).

\bibitem{tanzi2018}
\bibinfo{author}{Tanzi, L.} \emph{et~al.}
\newblock \bibinfo{title}{Feshbach resonances in potassium {Bose-Bose}
  mixtures}.
\newblock \emph{\bibinfo{journal}{Phys. Rev. A}} \textbf{\bibinfo{volume}{98}},
  \bibinfo{pages}{062712} (\bibinfo{year}{2018}).

\bibitem{cetina2015}
\bibinfo{author}{Cetina, M.} \emph{et~al.}
\newblock \bibinfo{title}{Decoherence of impurities in a {Fermi} sea of
  ultracold atoms}.
\newblock \emph{\bibinfo{journal}{Phys. Rev. Lett.}}
  \textbf{\bibinfo{volume}{115}}, \bibinfo{pages}{135302}
  (\bibinfo{year}{2015}).

\bibitem{fletcher2017}
\bibinfo{author}{Fletcher, R.~J.} \emph{et~al.}
\newblock \bibinfo{title}{Two-and three-body contacts in the unitary {Bose}
  gas}.
\newblock \emph{\bibinfo{journal}{Science}} \textbf{\bibinfo{volume}{355}},
  \bibinfo{pages}{377--380} (\bibinfo{year}{2017}).

\bibitem{braaten2010}
\bibinfo{author}{Braaten, E.}, \bibinfo{author}{Kang, D.} \&
  \bibinfo{author}{Platter, L.}
\newblock \bibinfo{title}{Short-time operator product expansion for rf
  spectroscopy of a strongly interacting {Fermi} gas}.
\newblock \emph{\bibinfo{journal}{Phys. Rev. Lett.}}
  \textbf{\bibinfo{volume}{104}}, \bibinfo{pages}{223004}
  (\bibinfo{year}{2010}).

\bibitem{Rath2013}
\bibinfo{author}{Rath, S.~P.} \& \bibinfo{author}{Schmidt, R.}
\newblock \bibinfo{title}{Field-theoretical study of the {Bose} polaron}.
\newblock \emph{\bibinfo{journal}{Phys. Rev. A}} \textbf{\bibinfo{volume}{88}},
  \bibinfo{pages}{053632} (\bibinfo{year}{2013}).

\bibitem{Knap2012}
\bibinfo{author}{Knap, M.} \emph{et~al.}
\newblock \bibinfo{title}{Time-dependent impurity in ultracold fermions:
  Orthogonality catastrophe and beyond}.
\newblock \emph{\bibinfo{journal}{Phys. Rev. X}} \textbf{\bibinfo{volume}{2}},
  \bibinfo{pages}{041020} (\bibinfo{year}{2012}).

\bibitem{Sommer2011}
\bibinfo{author}{Sommer, A.}, \bibinfo{author}{Ku, M.} \&
  \bibinfo{author}{Zwierlein, M.~W.}
\newblock \bibinfo{title}{Spin transport in polaronic and superfluid {Fermi}
  gases}.
\newblock \emph{\bibinfo{journal}{New Journal of Physics}}
  \textbf{\bibinfo{volume}{13}}, \bibinfo{pages}{055009}
  (\bibinfo{year}{2011}).

\bibitem{Bardon2014}
\bibinfo{author}{Bardon, A.~B.} \emph{et~al.}
\newblock \bibinfo{title}{Transverse demagnetization dynamics of a unitary
  {Fermi} gas}.
\newblock \emph{\bibinfo{journal}{Science}} \textbf{\bibinfo{volume}{344}},
  \bibinfo{pages}{722--724} (\bibinfo{year}{2014}).

\bibitem{Camacho-Guardian2018b}
\bibinfo{author}{Camacho-Guardian, A.} \& \bibinfo{author}{Bruun, G.~M.}
\newblock \bibinfo{title}{Landau effective interaction between quasiparticles
  in a {Bose-Einstein} condensate}.
\newblock \emph{\bibinfo{journal}{Phys. Rev. X}} \textbf{\bibinfo{volume}{8}},
  \bibinfo{pages}{031042} (\bibinfo{year}{2018}).

\bibitem{Alexandrov2010}
\bibinfo{author}{{Alexandrov}, A.~S.} \& \bibinfo{author}{{Devreese}, J.~T.}
\newblock \emph{\bibinfo{title}{{Advances in Polaron Physics}}}, vol.
  \bibinfo{volume}{159} (\bibinfo{publisher}{Springer-Verlag, Berlin},
  \bibinfo{year}{2010}).

\bibitem{Camacho-Guardian2018}
\bibinfo{author}{Camacho-Guardian, A.}, \bibinfo{author}{Pe\~na Ardila, L.~A.},
  \bibinfo{author}{Pohl, T.} \& \bibinfo{author}{Bruun, G.~M.}
\newblock \bibinfo{title}{Bipolarons in a {Bose-Einstein} condensate}.
\newblock \emph{\bibinfo{journal}{Phys. Rev. Lett.}}
  \textbf{\bibinfo{volume}{121}}, \bibinfo{pages}{013401}
  (\bibinfo{year}{2018}).

\bibitem{bloom2013}
\bibinfo{author}{Bloom, R.~S.}, \bibinfo{author}{Hu, M.-G.},
  \bibinfo{author}{Cumby, T.~D.} \& \bibinfo{author}{Jin, D.~S.}
\newblock \bibinfo{title}{Tests of universal three-body physics in an ultracold
  Bose-Fermi mixture}.
\newblock \emph{\bibinfo{journal}{Phys. Rev. Lett.}}
  \textbf{\bibinfo{volume}{111}}, \bibinfo{pages}{105301}
  (\bibinfo{year}{2013}).

\bibitem{wacker2016}
\bibinfo{author}{Wacker, L.~J.} \emph{et~al.}
\newblock \bibinfo{title}{Universal three-body physics in ultracold KRb
  mixtures}.
\newblock \emph{\bibinfo{journal}{Phys. Rev. Lett.}}
  \textbf{\bibinfo{volume}{117}}, \bibinfo{pages}{163201}
  (\bibinfo{year}{2016}).

\bibitem{zaccanti2009}
\bibinfo{author}{Zaccanti, M.} \emph{et~al.}
\newblock \bibinfo{title}{Observation of an Efimov spectrum in an atomic
  system}.
\newblock \emph{\bibinfo{journal}{Nat. Phys.}} \textbf{\bibinfo{volume}{5}},
  \bibinfo{pages}{586} (\bibinfo{year}{2009}).

\bibitem{roy2013}
\bibinfo{author}{Roy, S.} \emph{et~al.}
\newblock \bibinfo{title}{Test of the universality of the three-body Efimov
  parameter at narrow Feshbach resonances}.
\newblock \emph{\bibinfo{journal}{Phys. Rev. Lett.}}
  \textbf{\bibinfo{volume}{111}}, \bibinfo{pages}{053202}
  (\bibinfo{year}{2013}).

\bibitem{wacker2018}
\bibinfo{author}{Wacker, L.~J.} \emph{et~al.}
\newblock \bibinfo{title}{Temperature dependence of an Efimov resonance in
  $^{39}\mathrm{K}$}.
\newblock \emph{\bibinfo{journal}{Phys. Rev. A}} \textbf{\bibinfo{volume}{98}},
  \bibinfo{pages}{052706} (\bibinfo{year}{2018}).

\end{thebibliography}

\end{document}